Prevalence of algorithm-based qualitative (ABQ) method osteoporotic vertebral fracture in elderly Chinese men and women with reference to semi-quantitative (SQ) method: Mr. Os and Ms Os. (Hong Kong) studies

Xian Jun Zeng[1,4,§], Min Deng[1,§], Yì Xiáng J. Wáng[1], James F. Griffith[1], Lai Chang He[1,4], Anthony W. L. Kwok[3], Jason C. S. Leung[2], Timothy Kwok[2,4], and Ping Chung Leung[2]

Running title: osteoporotic vertebral fracture in elderly Chinese

X.-J. Zeng, M. Deng, Y.-X. J. Wang (*), J. F. Griffith, L-C He,

[1]Department of Imaging and Interventional Radiology, The Chinese University of Hong Kong, Prince of Wales Hospital, New Territories, Hong Kong SAR

*e-mail: yixiang_wang@cuhk.edu.hk  tel: (852) 2632 2289

J. C. Leung, T. Kwok, P. C. Leung.

[2]Jockey Club Centre for Osteoporosis Care and Control, School of Public Health and Primary Care, The Chinese University of Hong Kong, Prince of Wales Hospital, New Territories, Hong Kong SAR

Anthony W. L. Kwok,

[3]Department of Orthopedics and Traumatology, The Chinese University of Hong Kong, Prince of Wales Hospital, New Territories, Hong Kong SAR

T. Kwok.

[2,4]Department of Medicine and Therapeutics, The Chinese University of Hong Kong, Prince of Wales Hospital, New Territories, Hong Kong SAR

X.-J. Zeng, L-C He.

[4]Department of Radiology, The First Affiliated Hospital of Nanchang University, Nanchang, China

[§]: These two authors contributed equally to this study.

Prevalence of algorithm-based qualitative (ABQ) method osteoporotic vertebral fracture in elderly Chinese men and women with reference to semi-quantitative (SQ) method: Mr. Os and Ms Os. (Hong Kong) studies

Abstract

*Introduction:* This study evaluated algorithm-based qualitative (ABQ) method for vertebral fracture (VF) evaluation with reference to semi-quantitative (SQ) method and bone mineral density (BMD) measurement. *Methods:* Mr. OS (Hong Kong) and Ms. OS (Hong Kong) represent the first large-scale cohort studies on bone health in elderly Chinese men and women. The current study compared Genant's SQ method and ABQ method in these two cohorts. Based on quantitative measurement, the severity of ABQ method detected fractures was additionally classified into grade-1, grad-2, and grade-3 according to SQ's deformity criteria. The radiographs of 1,954 elderly Chinese men (mean: 72.3 years) and 1,953 elderly Chinese women (mean: 72.5 years) were evaluated. *Results:* according to ABQ, grade-1,-2,-3 VFs accounted for 1.89%, 1.74%, 2.25% in men, and 3.33%, 3.07%, and 5.53% in women. In men and women, 15.7% (35/223) and 34.5% (48/139) of vertebrae with SQ grade-1 deformity were ABQ(+, with fracture) respectively. In men and women, 89.7% (35/39) and 66.7% (48/72) of vertebrae with ABQ grade-1 fracture had SQ grade-1 deformity. For grade-1 change, SQ (-, negative without fracture) & ABQ (+, positive with vertebral cortex line fracture) subjects tend to have a lower BMD than the SQ(+)& ABQ(-) subjects. In subjects with SQ grade-2 deformity, those were also ABQ(+) tended to have a lower BMD than those were ABQ(-). In all grades, SQ(-)&ABQ(-) subjects tended to have highest BMD, while SQ(+)&ABQ(+)subjects tended to have lowest BMD. *Conclusion:* ABQ method may be more sensitive to VF associated mild lower BMD than SQ method.

Key words: Bone mineral density; Chinese; Epidemiology; Osteoporosis; Prevalence; Vertebral fractures.

**Introduction**

Vertebral fracture (VF) is the most common osteoporotic fracture. Prevalent VF predicts future osteoporotic fracture independently of bone mineral density (BMD) [1-4]. A vertebral compression fractures makes the diagnosis of osteoporosis independent of BMD level or ''T-score'' [5-9]. If a VF is present after the age of 50 year, the patient is at 5 times the risk of a future VF and double the risk of a hip fracture [10]. There are now effective bone protective and bone enhancing therapies, which for quite modest increases in bone mineral density (BMD) of 4%-12% reduce future VF risk by between 30% and 70% [11,12]. Although VFs cause only a modest proportion of the kyphosis that develops with increasing age, incident VFs are associated with progression of kyphosis, which in turn is associated with reduced pulmonary function, gastroesophageal reflux disease [13], reduced physical function [14], and possibly falls [15, 16]. A number of methods for diagnosing osteoporotic VF exist [17, 18]. The concordance across these methods still remains a matter of debate [19]. Quantitative morphometry (QM) uses ratios derived from direct vertebral body height measurements to define fractures [19 21]. Genant *et al* proposed semi-quantitative (SQ) visual grading which relies on subjective assessment of vertebral body height [22]. This SQ scoring system involves visual inspection of spinal radiographs by an experienced reader allowing exclusion of physiological variants in vertebral shape as well as non-fracture vertebral deformities which may be confused with vertebral fracture [22].

The algorithm-based qualitative (ABQ) method introduced a scheme to systematically rule out non-fracture deformity and diagnose osteoporotic VF [23]. The ABQ method, which particularly depends on the diagnosis of endplate fracture, was developed in an attempt to reduce the false-positive rate and the subjectivity associated with other diagnostic methods. This is achieved by incorporating specific criteria to identify osteoporotic fracture and to exclude non-osteoporotic deformity. Ferrar *et al.* [24] reported that inter-observer agreement for radiographic diagnosis of prevalent VF was better for the ABQ compared with the SQ method; and that agreement between ABQ and SQ was moderate. It was suggested that the utility of

ABQ method may be best for mild fractures where it can help differentiate true osteoporotic fractures from nonfracture deformities [25].

In Mr Os (Hong Kong) and Ms OS (Hong Kong) studies, 2000 Chinese elderly men and 2000 Chinese elderly women were studied to determine the relationship between anthropometric, lifestyle, medical and other factors with bone mineral density (BMD) at the hip and spine. Our results demonstrated that the age-specific VF prevalence of Chinese women is similar to other Japanese and Korean women and Latin American women [24]. This reinforces that the prevalence of VF tends to be similar across regions. Using the dataset of Mr. Os (Hong Kong) and Ms OS (Hong Kong), the purpose of this current study is three-fold: (1) to investigate the prevalence of VFs with ABQ method in elderly Chinese elderly men and women; (2) to compare the results of ABQ and SQ methods, and understand the reasons for disagreement between both methods; (3) to compare the bone mineral density (BMD) characteristics of men and women with and without VF according to ABQ method.

**Materials and Methods**

Two thousand Chinese men and two thousand Chinese women aged 65 or above were recruited from the local community by advertisements placed in housing estates and community centers for the elderly people for a prospective cohort study from August2001 to March 2003. The project was designed primarily to examine the BMD of older Chinese adults prospectively for 4 years.

All subjects were community dwelling, able to walk without assistance, without bilateral hip replacement and had the potential to survive the duration of the primary study as judged by their pre-existing medical status. Subjects were invited to the research center for interviews and physical examination. The recruit plan was designed so that the participants would represent the general elderly population in age and gender proportion. The study protocol was approved by the Chinese University of Hong Kong Ethics Committee. Written informed consent was obtained from all subjects. Data from the baseline evaluation were analyzed in the current report. BMD (g/cm2) at the total hip and spine (L1-L4) was measured by Hologic QDR-4,500 W densitometers (Hologic, Inc., Bedford, Mass. USA).

Left lateral thoracic and lumbar spine radiographs were obtained by adjusting exposure parameters according to participants' body weight and height. Hard copies of spine radiographs were taken and then digitalized later. Hard copy film was used for analysis primarily, aided with digitalized format for difficult cases using ImageJ software. The readers were blinded to clinical characteristics of the participants. The two readers (A & B) were both radiologists with more than ten years experience in reading spinal radiographs. Before the formal grading started, one month was given to allow the reader to familiarize themselves with the ABQ grading system, by comparing lateral lumbar spine radiographs from the Mr. OS (Hong Kong) and Ms. OS (Hong Kong) studies, as well as normal lumbar radiographs stored in our institution. Experience in SQ evaluation has been gained in previous study [24]. For all readings, the two readers read the images simultaneously, and consensus was reached by discussion. For both ABQ and SQ assessment, non-fractural changes of the vertebrae were evaluated according to radiological experience prior to morphometry measurement, which may be caused by deformities including developmental short vertebral height, cupid's bow deformity, Scheuermann disease, and Schmorl's nodes, degenerative remodeling [10, 18, 27-29]. The common developmental and/or acquired wedge deformities of the mid-thoracic and thoracic-lumbar regions, the reverse wedging of lower lumbar vertebrae and the common mild endplate bowing of the lower lumbar vertebrae were recognized. The SQ method as described initially does not require a radiological fracture sign as the pre-condition [22, 25, 30]. SQ grade-1 is usually termed as deformity, as it is known that some of the case may not represent true fracture.

Vertebrae T4 through L4 were evaluated by readers A and B using the ABQ method as described by Jiang *et al.* [30]. Each vertebra was classified to one of the following potential categories: (1) osteoporotic VF; (2) non-osteoporotic short vertebral height; (3) normal; (4) uncertain (possible osteoporotic fracture, but uncertain because of atypical appearances or poor image quality); or (5) unable to evaluate (poor image quality or not imaged). Osteoporotic

VF was identified when there was typical osteoporotic fracture of the central vertebral endplate, or fracture of the vertebral ring or cortex [30, 31]. Two modifications were introduced ABQ method. One is in addition to vertebral endplate fracture; we added fracture of any parts of the vertebral ring or cortex (vertebra cortex fracture, VCF). The second is that to facilitate follow-up and epidemiological studies, the severity of fracture identified by ABQ was determined by measurement based on reduction in vertebral height as follows: grade 1: 20-25% reduction in anterior, middle, and/or posterior height and a reduction of area 10-20%, grade 2, approximately 25-40% reduction in any height and a reduction in area 20-40%, grade 3, approximately 40% reduction in any height and area [22]. This approach is similar to the SQ grading scale except that there is no minimum threshold for reduction in vertebral height for ABQ definition of a prevalent fracture, whereas using the SQ approach, VF is diagnosed when vertebral height appears subjectively reduced by at least 20% compared to expected normal vertebral height at that particular level. The SQ method was diagnosed according to Genent's description [22]. As opposed to our previous study where SQ deformity or fracture was evaluated as the initially description by Genant et al [26], in the current study quantitative measurement was performed.

The intra-reader reproducibility *kappa* was tested to 0.78 for ABQ method, which was similar to the result of Ferrar *et al* [24]. For SQ grading, for the first step each vertebrae was radiologically assessed to exclude non-fractural deformity. In our previous study for SQ grading, we found the *kappa* for inter-reader agreement of 0.75 for SQ reading [26]. The main discord for inter-reader agreement disagreement related to the borderline cases in that a perceived reduction in vertebral height of close to 20% could be categorized as normal by one reader and mild vertebral fracture by another reader. Similarly, a perceived reduction in vertebral body height of close to 25% could be categorized as mild or moderate vertebral fracture. In this study consensus reading was adopted, we expected a better *kappa* value was achieved as actual measurement was taken instead of visual assessment of reduction in vertebral height.

Statistical analyses were performed using the statistical package SAS, version 9.1.3 (SAS Institute, Inc., Cary, NC, USA). Two sample independent t-tests were used for continuous variables while Chi-square tests were used for categorical variables. Logistic regression analysis was performed for significant factors. All statistical tests were two-sided. An α level of 5% was used as the level of significance.

**Results**

In general ABQ method requires a higher film quality than SQ method. During the analysis 46 (2.3%) spine radiographs for males and 47 (2.35%) spine radiographs for female (out of 2000 for each group) were adjudged to be of sub-optimal film quality for ABQ method assessment, leaving 1954 male subjects (mean age 72.3 years, range 65-92 years) and 1953 females (mean age 72.5 years, range 65-98 years) for final analysis. Suboptimal film quality included scoliosis subjects and films with insufficient X-ray exposure. There was no difference in age between the two sexes (P=0.417). None of these subjects' spines were diagnosed as having pathological fractures or diseases other than degenerative or osteoporotic change.

The prevalence of VF according to ABQ method and SQ method is presented in table 1. According to ABQ method, 115 men (5.9%, 95% CI: 4.8%-6.9%) and 233 women (11.9%, 95%CI: 10.5%-13.4%) had osteoporotic VF. In men, grade-1,-2,-3 VFs accounted for 1.89%, 1.74%, and 2.25% respectively, while in women grade-1, -2,-3 VFs accounted for 3.33%, 3.07%, and 5.53% respectively. The difference of VF prevalence between men and women was significant (p<0.001) with all grades of fracture being more prevalent in females. ABQ VF shows a lower prevalence than SQ VF for all grades of fracture (table1). The highest SQ positive/ABQ negative subject occurred in SQ grade 1 deformity, particularly for males (table 1).

Prevalence of VF at three age groups (65~69 yrs, 70~79 yrs,≥80 yrs) is shown in table 2 and visually in Fig 1A, the prevalence of VF was closely related to age. Prevalence of VF according to spine and hip BMD status is shown in table 3 and visually in Fig 1B. The prevalence of VF increased as BMD decreased. In particular there was a sharp increase of VF related to osteoporosis defined by hip BMD.

Spine BMD and hip BMD according to VF are shown in table 4 and visually with Supplementary Fig 1. As expected both spine BMD and hip BMD decreased in subjects with higher grade VF. For hip BMD, the notable difference was shown in SQ grade 1 deformity vs. ABQ grade 1 VF, with ABQ grade 1 VF having a lower mean BMD than SQ grade 1 deformity, thus showing that ABQ grade 1 VF is more closely to be associated with low BMD than SQ grade 1 VF. However, as expected, similar BMD was shown for SQ grade 2/3 VF and ABQ grade 2/3 VF (Supplementary Fig 1).

The discordance of ABQ vs. SQ grading is shown in Supplementary table 1. In men the majority (84.3%, 188/223) of the vertebrae graded as SQ grade-1 deformity were ABQ (-). In women, 91(65.5%) of 139 vertebrae with SQ grade-1 deformity were ABQ(-). In other words, females with a SQ grade-1 deformity were more likely to have VCF than males were. In men, 89.7% (35/39) of vertebrae with ABQ grade-1 fracture also had SQ grade-1 deformity. In women 66.7% (48/72) of vertebrae with ABQ grade-1 VF also had SQ grade-1 VF.

The BMD values in VF positive subjects using either the SQ method or the ABQ are shown in Table 5. In all grades, both SQ and ABQ VF negative subjects tended to have the highest BMD, while SQ and ABQ VF positive subjects tend to have lowest BMD. For grade-1 VF, SQ(-) & ABQ(+) subjects tend to have a higher BMD than the SQ(+)& ABQ(-) subjects. In subjects with SQ grade-2 deformity, those were also ABQ positive tended to have a lower BMD than those was ABQ negative. The location of VF prevalence is highest at T12 and L1, second highest at T11 and L2. A higher VF rate is also seen at T8 in women, but less so in men (Supplementary Figure 2). Counts on exact vertebral height reductions in ABQ VFs are shown in supplementary Figure 3&4.

**Discussion**

Either symptomatic (painful) or asymptomatic (radiographically defined) VF has clinical implications if unrecognized and untreated. Radiographic VF can be asymptomatic [1-6]. Either type of fractures reduces pulmonary vital capacity, leads to a greater risk of other fragility fractures at both vertebral bodies and other skeletal sites. Patients can have normal T-scores and yet also have VF where the bone strength is impaired by poor bone quality, especially in diabetes mellitus and chronic kidney disease subjects. However, the best approach to diagnose remains to be established [17, 31].

Difficulties remain with correctly classifying vertebrae whose height reduction does not reach 20% though with features of fracture or an apparent height reduction of more than 20% but no additional features of fracture. Vertebral height may also appear decreased as a result of image obliquity, diseases such as Scheurermann's disease and physiological wedging that can mimic vertebral fracture [10, 18, 25, 29, 31]. The upper lumbar vertebrae are often physiologically wedged at the transition between lumbar lordosis and thoracic kyphosis as are the mid-thoracic vertebrae to a mild degree making VF diagnosis sometimes difficult at this level. Apparent reduction in vertebral height without endplate or other sign of cortex fracture was categorized by ABQ method as non-osteoporotic short vertebral height, which is not significantly associated with low bone density, nor is it significantly associated with incident vertebral fracture identified on VF [33].

Ferrar *et al.* [32] observed a strong association between prevalent fractures identified by ABQ and the incidence of new vertebral fractures, even after adjustment for age and BMD. Jiang *et al* [23] found *kappa* statistics between 0.39 and 0.64 comparing ABQ with the quantitative morphometry method. ABQ compared with SQ yielded kappa statistics of 0.30 to 0.58 [24]. Ferrar *et al.* [24] noted that inter-observer agreement for radiographic diagnosis of prevalent VF was significantly better for the ABQ compared with the SQ method; and agreement between ABQ and SQ was moderate. Of all methods, ABQ readings resulted in the lowest recordings of vertebral fracture prevalence [24]. Ferrar *et al.* [32] also observed low BMD at the lumbar spine in women with mild ABQ but not SQ fractures. However, ABQ method has not been very well validated except the Sheffield group.

The current study compared ABQ method and SQ method in a large Chinese cohort. In this study, we introduced two additionally approaches with the originally described ABQ method and SQ method. The first is we quantitatively measured vertebral deformity criteria for SQ, i.e grade 1: 20-25% reduction in anterior, middle, and/or posterior height and a reduction of area 10-20%, grade 2, approximately 25-40% reduction in any height and a reduction in area 20-40%, grade 3, approximately 40% reduction in any height and area. The original description by Genant *et al* did not require quantitative measurement [22], but as described by Jiang *et al* [23] and also our own experience that visual estimation of reduction in vertebral height or area is difficult to achieve accurately without the aid of direct measurements. Visual estimation of reduction in vertebral height or area may be applicable for clinical diagnostic purpose, but it is difficult to achievement sufficient consistency for epidemiological studies and follow-up studies. Therefore for ABQ method detected VCF we introduced grade-1, 2, 3, we feel this will be useful for epidemiological studies and longitudinal follow-up studies. Secondly, we did not limit vertebral cortex fracture (VCF) to the endplate. It is possible that vertebra compress and crush can lead to fracture of the anterior cortex or/and posterior cortex fracture, but without endplate. Therefore the term VCF is used in this study. In this study, the prevalence of ABQ fracture increased as the subject age increased. ABQ fracture was more common in osteoporotic subjects than osteopenic subjects and least common in subjects with normal BMD. The BMD of ABQ grade-1 VF subjects was lower than that of ABQ(-) subjects. In contrast, subjects with SQ grade-1 deformity have a similar BMD to subjects without fracture. These results show the relevance of ABQ to lower BMD in the subjects, and the relationship between VF prevalence and low BMD seems to be stronger with the ABQ method than with the SQ method. Ferrar *et al.* [24] also found that the association between low BMD and mild VF were stronger by ABQ method than by SQ method. Consistent with previous publications, our study showed a lower VF prevalence when evaluated by ABQ than by SQ, as many SQ grade=1 deformity had no vertebral cortex fracture (VCF).

In this study a number of SQ grade-2 deformities did not have VCF. This is conceivable when the deformity magnitude was at low end of the grade-2 spectrum, i.e. a little over 20% decrease in vertebral height. Szulc et al. [30] suggested for male thoracic vertebrae, SQ grade-2 VF should be increased to 30%-40% rather than the current criteria of >25%-40%. In some other cases, scoliosis might have influenced so that vertebral endplates could not be reliably assessed. Our results showed all SQ grade-3 deformities had VCF. Our data also showed it is also possible for a vertebra to have ABQ grade-1 fracture but without SQ grade-1 deformity VF. This accounted for 10.3% (4/39) of SQ grade-1 deformity in males and 33.3 (24/72) SQ grade-1 deformity in females.

There has been debate how such as signs of fracture, including lack of end plate parallelism, end plate depression, buckling of cortical margin, and loss of vertical continuity with adjacent vertebrae, should be incorporated into the diagnosis of SQ fracture [25]. If the signs are strictly required for diagnosis, then the difference between SQ and ABQ methods would merge. In this study the authors chose to use the initial definition of SQ method which has been popularly used [22, 25]. Our results differ in some aspects from the reports of authors in Sheffield. In a smaller sample, Jiang et al reported the prevalence of VF in postmenopausal women was 7% with ABQ method and 24% SQ methods [23]; while in our study the prevalence of VF in postmenopausal women was 11.93% with ABQ method and 16.08% with SQ method. This could be due to two reasons, firstly our subjects were older (mean age 72.5 years, range 65-98 years) than Jiang et al's subjects (mean 64.4 years, range: 50 to 85 years), therefore higher ABQ VF rate in our study. Radiological evaluation was used to rule out non-fractural deformity during SQ evaluation, therefore lower SQ VF rate in our study. However, while Ferrar et al [24] reported the prevalence of VF in elderly men was 7% with ABQ method and 24% SQ methods (≥65 yr); in our study the prevalence of VF in elderly men was 5.88% with ABQ method and only 13.2 % with SQ method (mean age 72.3 years, range 65-92 years). Therefore in our study with ABQ method, elderly men have much lower VF rate than women (5.88% vs. 11.93%), while in

Jiang *et al* and Ferrar *et al*'s data it was 7% for elderly women and 10% for elderly men (23, 24). However, Jiang *et al*'s data and Ferrar *et al*'s data were not age-matched.

In our study with SQ method, elderly men had lower VF rate than women (13.2 % vs. 16.08 %). The prevalence of VF in men has not been as intensively studied as in women in literature. In Tromsø Study of Norway, with vertebral morphometry method Waterloo *et al* [33] reported a VF of 14.0% (165/1177) for elderly men (age: 65.4 ±8.8 years) and 12.2% (197/1418) for elderly women (age: 65.7 ±8.4 years), with slightly higher VF rate in elderly men.

In MrOS Sweden study, using SQ method Karlsson *et al.* [34] reported 15.1% (215/1427) VF prevalence for elderly men. Using SQ method Kucukler *et al.* [35] reported 17.6% VF prevalence in small sample of elderly males (age: 74.4 ± 0.7 years). Our VF prevalence in men had slightly lower rate compared with these reports. It has been reported that Asians have slight lower spine VF rate than Caucasians [26]. However, Chinese, Japanese, and South Korean have similar spine VF rat [26]. The skeletal location of VF in this study was highest at thoracolumbar region which agrees with Van der Klift *et al*'s report [36]. The middle thoracic spine VF occurs more often at T8, because thoracic kyphosis increases the axial load on the vertebral end plates this region [37]. However, the distribution of VF fractures in this study also differs from the Sheffield data [23], with less middle thoracic VF in our results. Though ethnicity factors might have caused these differences, but it is also possible that our SQ evaluation excluded more physiological wedge-shaped thoracic vertebrae [30]. ABQ method has been suggested to be a more accurate method of assessing prevalent VF and could reduce the false-positive rate and produce a more accurate evaluation of a patient's future fracture risk which this is a very relevant consideration for epidemiology and clinical follow-up studies. However, ABQ has been not well studied beside the reports by Jiang et al and Ferrar *et al* [22, 23].

This study established the ABQ determined VF prevalence in elderly Chinese population, and demonstrated the closer association of ABQ determined VF with lower BMD than SQ determined VF, particularly for milder grades. The disadvantages of ABQ method, according to the experiences of authors of this study but not fully documented, include it is time-consuming and expertise-dependant. Future evaluation of the concordance between various proposed methods will allow one to establish their benefits and limitations, and most importantly, optimize their effectiveness for use in epidemiology and clinical follow-up studies. At this moment, our results seem to recommend for search studies we need to look at both vertebral cortex fracture sign and quantify the extent of vertebra height or area reduction.

**Acknowledgments:**

This study is partially funded by National Institute of Health R01 Grant AR049439 -01A1and the Research Grants Council Earmarked Grant CUHK 4101/02M.

**Conflicts of interest:**

Yì Xiáng Wáng, Xian Jun Zeng, Min Deng, James F. Griffith, Lai Chang He, Anthony W. L. Kwok, Jason C. S. Leung, Timothy Kwok and Ping Chung Leung declare that they have no conflict of interest.

Table 1, Comparison of semi-quantitative (SQ) vs. algorithm-based qualitative (ABQ) in assessing spine osteoporotic vertebral fracture in Mr. OS (Hong Kong) and Ms. OS (Hong Kong) studies.

|  | Men | | Women | | p-value of chi square | |
|---|---|---|---|---|---|---|
|  | SQ (n=1954) | ABQ (n=1954) | SQ (n=1953) | ABQ (n=1953) | SQ (M vs F) | ABQ (M vs F) |
| Grade 0 | 86.80% | 94.12% | 83.92% | 88.07% |  |  |
| Grade 1 | 8.34% | 1.89% | 5.07% | 3.33% | 0.0004 | 0.0022 |
| Grade 2 | 2.61% | 1.74% | 5.12% | 3.07% | <.0001 | 0.0030 |
| Grade 3 | 2.25% | 2.25% | 5.89% | 5.53% | <.0001 | <.0001 |
| Total (grade1-3) | 13.20% | 5.88% | 16.08% | 11.93% | 0.0111 | <.0001 |

Grade 1, 2, 3 indicate prevalence of fracture in each grades.

Table 2, semi-quantitative (SQ) and algorithm-based qualitative (ABQ) evaluated osteoporotic vertebral fracture prevalence in three age groups.

|  | 65~69 (yrs) | 70~79 (yrs) | ≥80 (yrs) | P-value for association |
|---|---|---|---|---|
| Men (SQ grade 1, 2,3) | 10.2% (66/ 650) | 13.4% (151/ 1128) | 23.3% (41/ 176) | <.0001 |
| Men (SQ grade 2,3) | 2.5% (16/ 650) | 5.2% (59/ 1128) | 11.4% (20/ 176) | <.0001 |
| Women (SQ grade 1, 2,3) | 10.3% (68/ 662) | 16.6% (181/ 1089)* | 32.2% (65/ 202)* | <.0001 |
| Women (SQ grade 2,3) | 5.1% (34/ 662)* | 12.6% (137/ 1089)* | 21.8% (44/ 202)* | <.0001 |
| Men (ABQ grade 1, 2,3) | 3.4% (22/ 650) | 6.3% (71/ 1128) | 12.5% (22/ 176) | <.0001 |
| Women (ABQ grade 1, 2,3) | 6.0% (40/ 662)* | 13.1% (143/ 1089)* | 24.8% (50/ 202)* | <.0001 |

\* *p*-value <0.05, comparing men with women.

Table 3, Prevalence semi-quantitative (SQ) and algorithm-based qualitative (ABQ) evaluated osteoporotic vertebral fracture prevalence among normal BMD, osteopenia and osteoporosis subjects.

|  | According to Spine BMD | | | | According to Hip BMD | | | |
|---|---|---|---|---|---|---|---|---|
|  | **normal** | **osteopenia** | **osteoporosis** | **P-value for association** | **normal** | **osteopenia** | **osteoporosis** | **P-value for association** |
| Men (SQ grade 1, 2,3) | 11.0% (129/1169) | 15.5% (88/568) | 16.2% (31/191) | 0.0047 | 10.8% (103/950) | 14.1% (126/896) | 26.9% (29/108) | <.0001 |
| Men (SQ grade 2,3) | 3.3% (38/1169) | 6.9% (39/568) | 7.3% (14/191) | 0.0004 | 3.1% (29/950) | 5.1% (46/896) | 18.5% (20/108) | <.0001 |
| Women (SQ grade 1, 2,3) | 11.2% (42/374)* | 12.3% (88/715) | 21.1% (177/838)* | <.0001 | 8.6% (46/535) | 14.6% (145/994)* | 29.0% (123/424) | <.0001 |
| Women (SQ grade 2,3) | 6.4% (24/374)* | 8.5% (61/715) | 15.3% (128/838)* | <.0001 | 4.7% (25/535) | 9.3% (92/994)* | 23.1% (98/424) | <.0001 |
| Men (ABQ grade 1, 2,3) | 3.9% (45/1169) | 8.3% (47/568) | 8.9% (17/191) | <.0001 | 3.5% (33/950) | 6.9% (62/896) | 18.5% (20/108) | <.0001 |
| Women(ABQ grade 1, 2,3) | 8.6% (32/374)* | 8.5% (61/715) | 16.0% (134/838)* | <.0001 | 7.1% (38/535)* | 10.1% (100/994)* | 22.4% (95/424) | <.0001 |

Note: * *p*-value <0.05, comparing men with women.

Table 4, Spine BMD and hip BMD (g/cm$^2$) according to SQ and ABQ evaluations of the Mr. OS (Hong Kong) and Ms. OS (Hong Kong) study subjects

|  | Spine BMD | | Hip BMD | |
|---|---|---|---|---|
|  | **Men** | **Women** | **Men** | **Women** |
| SQ method | | | | |
| grade-0 | 0.954±0.179 | 0.761±0.147 | 0.870±0.126 | 0.720±0.115 |
| grade-1 | 0.943±0.174 | 0.744±0.154 | 0.856±0.125 | 0.696±0.100 |
| grade-2 | 0.905±0.173 | 0.700±0.150 [0] | 0.822±0.117 [0] | 0.642±0.120 [0,1] |
| grade-3 | 0.824±0.205 [0,1] | 0.681±0.144 [0,1] | 0.745±0.152 [0,1,2] | 0.635±0.108 [0,1] |
| ABQ method | | | | |
| grade-0 | 0.954±0.179 | 0.759±0.146 | 0.869±0.126 | 0.718±0.114 |
| grade-1 | 0.892±0.162 | 0.740±0.176 | 0.819±0.103 | 0.697±0.134 |
| grade-2 | 0.905±0.163 | 0.699±0.155 [0] | 0.817±0.098 | 0.646±0.127 [0] |
| grade-3 | 0.824±0.205 [0] | 0.680±0.147 [0] | 0.745±0.152 [0,1] | 0.630±0.105 [0,1] |

[0] p-value<0.05 (Bonferroni adjusted), comparing grade 1, 2 or 3 with grade 0
[1] p-value<0.05 (Bonferroni adjusted), comparing grade 2 or 3 with grade 1
[2] p-value<0.05 (Bonferroni adjusted), comparing grade 3 with grade 2

Table 5. BMD values in different semi-quantitative (SQ) method and algorithm-based qualitative (ABQ) method negative and/or positive groups.

| Men | | SQ (-) & ABQ (-) [a] | SQ(+)& ABQ(-) [b] | SQ(-)& ABQ(+) [c] | SQ(+)& ABQ(+) [d] |
|---|---|---|---|---|---|
| Spine BMD | Grade 0 | 0.954±0.179 | | | |
| | Grade 1 | | 0.956±0.176 | 1.052±0.109 | 0.874±0.151 [a,b] |
| | Grade 2 | | 0.907±0.197 | - | 0.857±0.201 [a] |
| | Grade 3 | | - | - | 0.824±0.205 [a] |
| Hip BMD | Grade 0 | 0.870±0.126 | | | |
| | Grade 1 | | 0.866±0.128 | 0.871±0.041 | 0.809±0.111 [a,b] |
| | Grade 2 | | 0.831±0.151 | - | 0.777±0.138 [a] |
| | Grade 3 | | - | - | 0.745±0.152 [a] |
| Women | | SQ (-) & ABQ (-) [a] | SQ(+)& ABQ(-) [b] | SQ(-)& ABQ(+) [c] | SQ(+)& ABQ(+) [d] |
| Spine BMD | Grade 0 | 0.761±0.146 | | | |
| | Grade 1 | | 0.727±0.144 [a] | 0.717±0.161 | 0.702±0.182 [a] |
| | Grade 2 | | 0.703±0.144 [a] | - | 0.675±0.155 [a] |
| | Grade 3 | | 0.711±0.074 | - | 0.680±0.147 [a] |
| Hip BMD | Grade 0 | 0.721±0.114 | | | |
| | Grade 1 | | 0.674±0.098 [a] | 0.683±0.148 | 0.654±0.130 [a] |
| | Grade 2 | | 0.637±0.111 [a] | - | 0.627±0.128 [a] |
| | Grade 3 | | 0.713±0.133 | - | 0.630±0.105 [a] |

[a] $p<0.05$ comparing [b], [c] or [d] with [a]

[b] $p<0.05$ comparing [c] or [d] with [b]

Supplementary Table 1. The discordance of ABQ vs. SQ evaluations

|  | Men | | Women | |
|---|---|---|---|---|
|  | SQ (n=1954) | ABQ (n=1954) | SQ (n=1953) | ABQ (n=1953) |
| Grade 1 | 8.34% (163; 223*)<br>(abq+) =1.74%(34;35*)<br>(abq-)=6.60%(129;188*) | 1.89%(37; 39*)<br>(sq+)=1.74%(34;35*)<br>(sq-)=0.15% (3;4*) | 5.07%(99;139 *)<br>(abq+)=2.15%(42;48*)<br>(abq-)=3.48%(57;91*) | 3.33% (65;72*)<br>(sq+)=2.15%(42, 48*)<br>(sq-)=1.18%(23, 24*) |
| Grade 2 | 2.61%(51;57*)<br>(abq+)=1.74%(34;35*)<br>(abq-)=0.87%(17;22*) | 1.74% (34;35*)<br>(sq+)=1.74% (34;35*) | 5.12%(100;125*)<br>(abq+)=3.07%(60;74*)<br>(abq-)=2.05%(40;51*) | 3.07%(60;74 *)<br>(sq+)=3.07%(60; 74*) |
| Grade 3 | 2.25%(44;53*)<br>(abq+)=2.25%(44;53*) | 2.25%(44;53*)<br>(sq+)=2.25%(44;53*) | 5.89%(115; 149*)<br>(abq+)=5.53%(108;142*)<br>(abq-)=0.36%(7;7*) | 5.53%(108;142*)<br>(sq+)=5.53%(108;142*) |
| Total (grade1-3) | 13.20%(258; 333*)<br>(abq+)=5.73%(112;123*)<br>(abq-)=7.47%(146;210*) | 5.89%(115;127*)<br>(sq+)=5.73%(112;123*)<br>(sq-)=0.15%(3;4*) | 16.08%(314;413*)<br>(abq+)=10.29%(201;264*)<br>(abq-)=5.79%(113;149*) | 11.93%(233; 288*)<br>(sq+)=10.75%(210;264*)<br>(sq-)=1.18%(23;24*) |

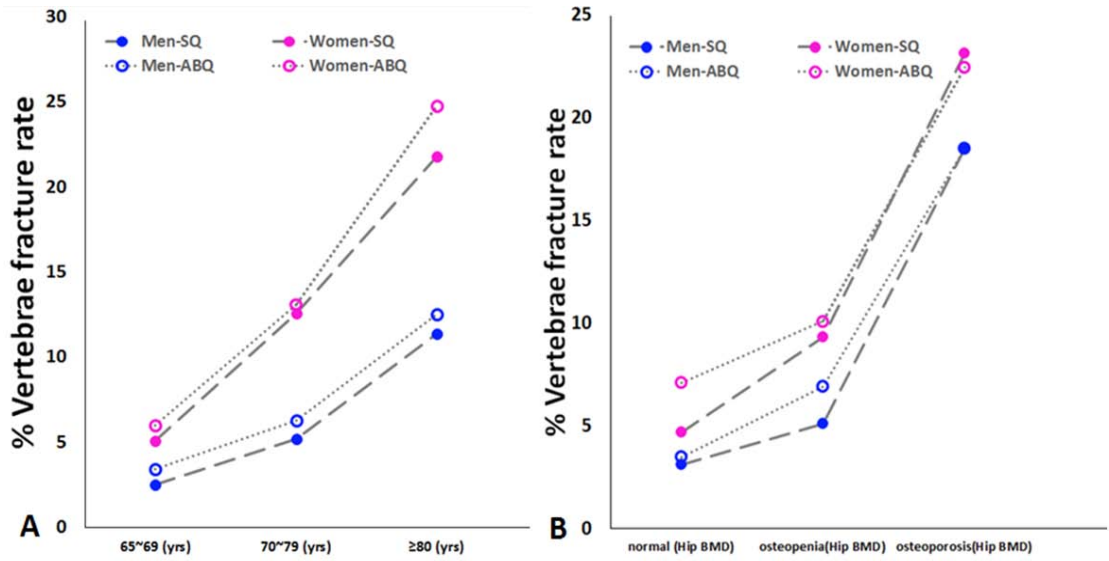

Fig 1, Prevalence of osteoporotic vertebral fracture among three age groups (65~69 yrs, 70~79 yrs, and ≥ yrs; A), and among normal BMD, osteopenia and osteoporosis subjects (B). Semi-quantitative (SQ) included deformities grade-2 and 3 only; algorithm-based qualitative (ABQ) included fracture grade-1, 2 and 3.

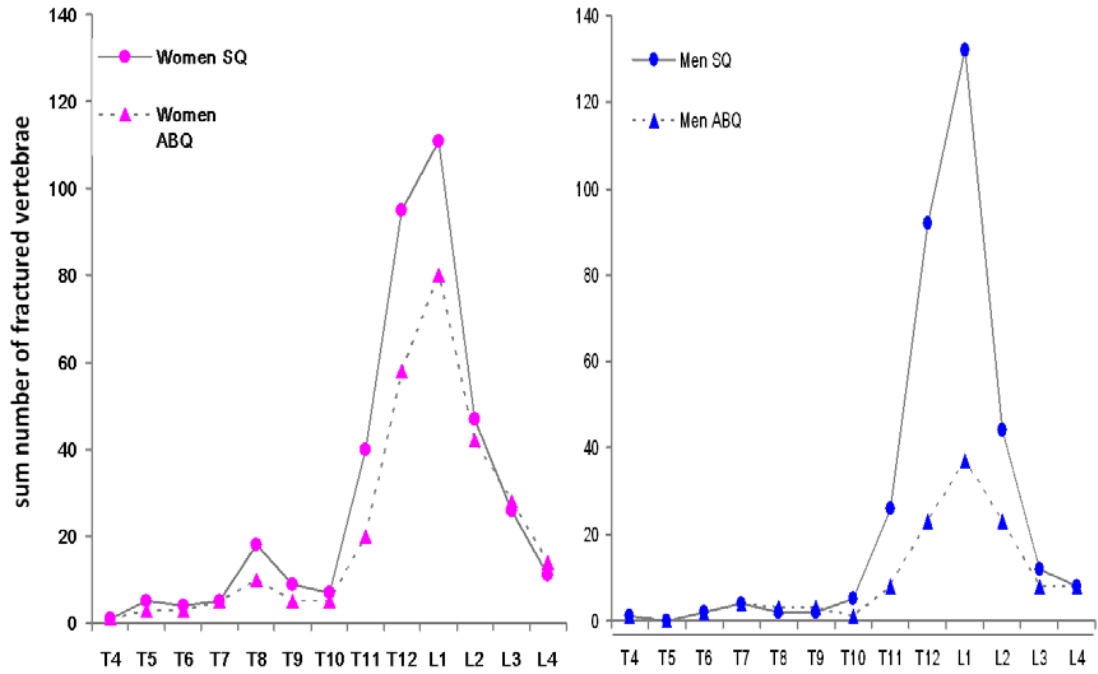

Supplementary Fig 2: The location of osteoporotic fracture distribution. It is highest at T12 and L1, second highest at T11 and L2. A higher VF rate is also seen at T8 in women, but less so in men.

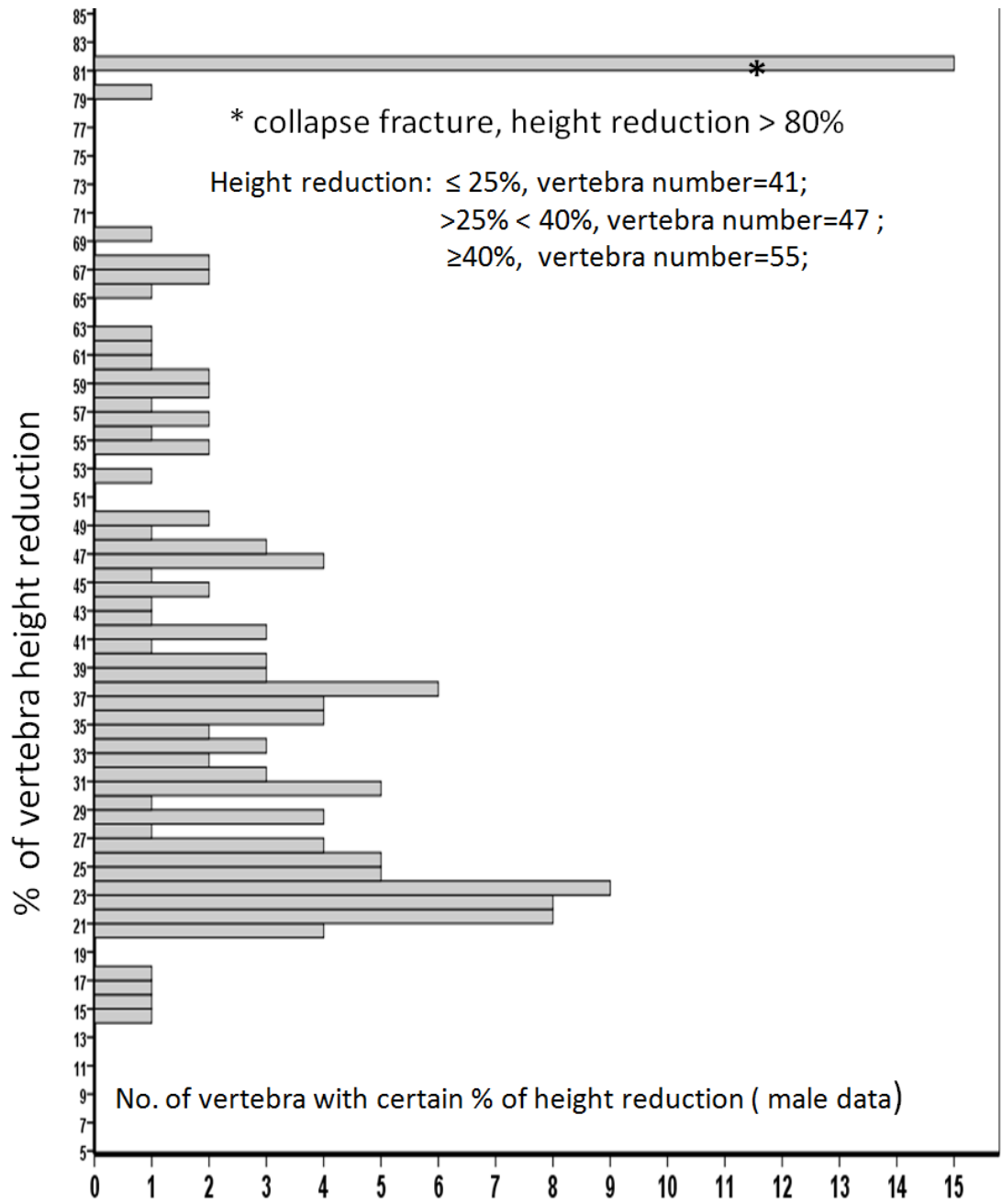

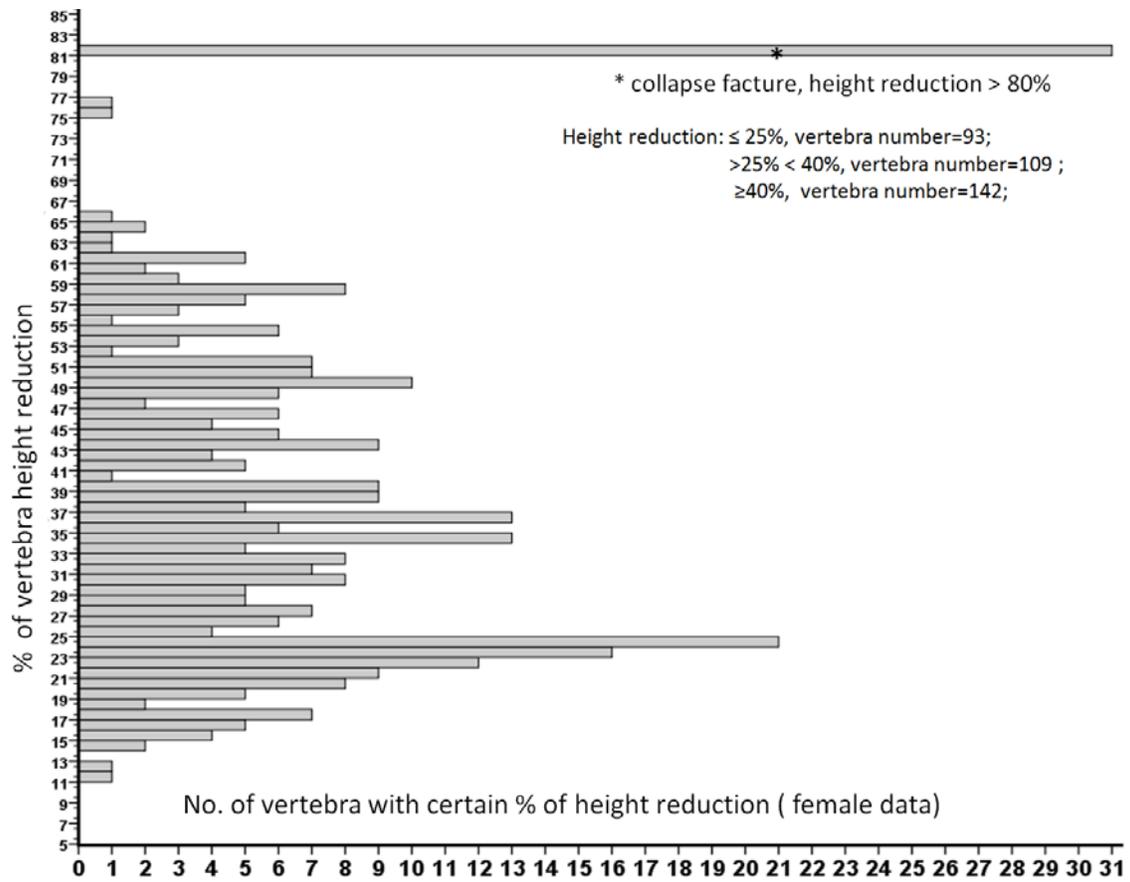